\pdfoutput=1

\documentclass[10pt,twocolumn,prd,aps,amssymb,amsmath,tightenlines,showpacs]{revtex4}

\usepackage{graphicx}

\newcommand{\cC}{\ensuremath{\mathcal{C}}}
\newcommand{\cB}{\ensuremath{\mathcal{B}}}
\newcommand{\cQ}{\ensuremath{\mathcal{Q}}}
\newcommand{\cP}{\ensuremath{\mathcal{P}}}
\newcommand{\cT}{\ensuremath{\mathcal{T}}}

\newcommand{\half}{\mbox{$\textstyle{\frac{1}{2}}$}}

\begin{document}

\title{Matrix representation of the time operator}

\author{Carl M. Bender$^a$}\email{cmb@wustl.edu}
\author{Mariagiovanna Gianfreda$^b$}\email{Maria.Gianfreda@le.infn.it}
\affiliation{$^a$Department of Physics, Kings College London, Strand, London
WC2R 1LS, UK\footnote{Permanent address: Department of Physics, Washington
University, St. Louis, MO 63130, USA.}\\
$^b$Dipartimento di Fisica, Universit\`a del Salento and
I.N.F.N. Sezione di Lecce, Via Arnesano, I-73100 Lecce, Italy}

\date{\today}

\begin{abstract}
In quantum mechanics the time operator $\Theta$ satisfies the commutation
relation $[\Theta,H]=i$, and thus it may be thought of as being canonically
conjugate to the Hamiltonian $H$. The time operator associated with a given
Hamiltonian $H$ is not unique because one can replace $\Theta$ by $\Theta+
\Theta_{\rm hom}$, where $\Theta_{\rm hom}$ satisfies the homogeneous condition
$[\Theta_{\rm hom},H]=0$. To study this nonuniqueness the matrix elements of
$\Theta$ for the harmonic-oscillator Hamiltonian are calculated in the
eigenstate basis. This calculation requires the summation of divergent series,
and the summation is accomplished by using zeta-summation techniques. It is
shown that by including appropriate homogeneous contributions, the matrix
elements of $\Theta$ simplify dramatically. However, it is still not clear
whether there is an optimally simple representation of the time operator.
\end{abstract}


\maketitle

\section{Nonuniqueness of the Time Operator for the Quantum Harmonic Oscillator}
\label{s1}
The {\it time operator} $\Theta$ in quantum mechanics satisfies the commutation
relation
\begin{equation}
\left[\Theta,H\right]=i,
\label{e1}
\end{equation}
where $H$ is the Hamiltonian \cite{S1,S2}. Even though time is not a dynamical
variable, one may regard $\Theta$ as being a time operator because if one allows
a normalized state $|A\rangle$ at $t=0$ to evolve to the state $|B\rangle=e^{-i
Ht}|A\rangle$ at time $t$, then as a consequence of (\ref{e1}),
\begin{equation}
\langle B|\Theta|B\rangle-\langle A|\Theta|A\rangle=t.
\label{e2}
\end{equation}

It is not known how to solve (\ref{e1}) for the time operator $\Theta$ for an
arbitrary Hamiltonian $H$. Even for the comparatively simple case of the
quantum-harmonic-oscillator Hamiltonian
\begin{equation}
H=\half p^2+\half q^2
\label{e3}
\end{equation}
the calculation of $\Theta$ is difficult because it involves singular operators
such as $1/p$ and $1/q$. The time operator for the quantum harmonic oscillator
was studied in detail at a formal level in Ref.~\cite{S3}. The approach taken in
\cite{S3} was to express $\Theta$ as a series of the form
\begin{equation}
\Theta=\sum_{m,n}a_{m,n}T_{m,n},
\label{e4}
\end{equation}
where $a_{m,n}$ are numerical coefficients and $T_{m,n}$ (for integer $m$ and
$n$) are a set of Hermitian basis operators.

For $m,n\geq0$ the operator $T_{m,n}$ is defined as a totally symmetric average
over all possible orderings of $m$ factors of $p$ and $n$ factors of $q$. For
example,
\begin{eqnarray}
T_{0,0}&=&1,\nonumber\\
T_{1,0}&=&p,\nonumber\\
T_{1,1}&=&\half(pq+qp),\nonumber\\
T_{1,2}&=&\textstyle{\frac{1}{3}}(pqq+qpq+qqp),\nonumber\\
T_{2,2}&=&\textstyle{\frac{1}{6}}(ppqq+qqpp+pqqp+qppq+qpqp+pqpq),\nonumber
\end{eqnarray}
and so on. The operator $T_{m,n}$ is the quantum-mechanical generalization of
the classical product $p^mq^n$.

It is easy to evaluate commutators and anticommutators of the operators
$T_{m,n}$. For example, as shown in Ref.~\cite{S4}, the operators $T_{m,n}$ obey
simple commutation and anticommutation relations:
\begin{eqnarray}
\left[p,T_{m,n}\right]&=&-inT_{m,n-1},\nonumber\\
\left[q,T_{m,n}\right]&=&imT_{m-1,n},\nonumber\\
\left\{p,T_{m,n}\right\}&=&2T_{m+1,n},\nonumber\\
\left\{q,T_{m,n}\right\}&=&2T_{m,n+1}.
\label{e5}
\end{eqnarray}
By combining and iterating the results in (\ref{e5}), one can establish
additional useful commutation relations for $T_{m,n}$:
\begin{eqnarray}
\left[p^2,T_{m,n}\right]&=&-2inT_{m+1,n-1},\nonumber\\
\left[q^2,T_{m,n}\right]&=&2imT_{m-1,n+1}.
\label{e6}
\end{eqnarray}

The totally symmetric operators $T_{m,n}$ can be re-expressed in Weyl-ordered
form \cite{S4}:
\begin{eqnarray}
T_{m,n}&=&\frac{1}{2^m}\sum_{k=0}^m\binom{m}{k}p^kq^np^{m-k}\nonumber\\
&=&\frac{1}{2^n}\sum_{k=0}^n\binom{n}{k}q^k p^m q^{n-k},
\label{e7}
\end{eqnarray}
where $m,\,n=0,\,1,\,2,\,3,\,\cdots$. The proof that $T_{m,n}$ equals the
binomial-summation Weyl-ordered forms above is inductive and requires repeated
use of the Heisenberg algebraic property that $[q,p]=i$. The advantage of
introducing the Weyl-ordered form of $T_{m,n}$ is that it allows one to extend
the totally symmetric operators $T_{m,n}$ either to negative values of $n$ by
using the first of these formulas or to negative values of $m$ by using the
second of these formulas. The commutation and anticommutation relations in
(\ref{e5}) and (\ref{e6}) remain valid when $m$ is negative or when $n$ is
negative.

In Ref.~\cite{S3} it was shown that for the harmonic-oscillator Hamiltonian
(\ref{e3}) the commutation relations in (\ref{e6}) may be used to obtain an
exactly solvable recursion relation for the coefficients $a_{m,n}$ in
(\ref{e4}). The exact solution for $\Theta$ has the series representation
\begin{equation}
\Theta_{\rm min}=\sum_{k=0}^\infty\frac{(-1)^{k+1}}{2k+1}T_{2k+1,-2k-1}.
\label{e8}
\end{equation}
In Ref.~\cite{S3} the solution in (\ref{e8}) was called the {\em minimal}
solution because of its relative simplicity.

We emphasize here that $\Theta_{\rm min}$ is merely a {\em particular} solution
to the inhomogeneous linear equation (\ref{e1}). To find other solutions to
(\ref{e1}), we combine $\Theta_{\rm min}$ with solutions to the associated
homogeneous equation
\begin{equation}
\left[H,\Theta_{\rm hom}\right]=0.
\label{e9}
\end{equation}
The subscript emphasizes that $\Theta_{\rm hom}$ is a solution to the {\em
homogeneous} equation (\ref{e9}).

In Ref.~\cite{S3} it was noted that the solution to (\ref{e1}) is not unique and
it was claimed that any two operators satisfying this commutation relation must
differ by a function of the Hamiltonian $H$ because this difference commutes
with $H$. (We will see in Sec.~\ref{s2} that this claim was actually not
correct.)

One can interpret (\ref{e8}) by going to the classical limit $T_{m,n}\to p^m
q^n$. In this limit the series (\ref{e8}) reduces to
\begin{equation}
\Theta_{\rm classical}=\sum_{k=0}^\infty\frac{(-1)^{k+1}}{2k+1}p^{2k+1}q^{-2k-1}
={\rm arctan}\frac{p}{q}.
\label{e10}
\end{equation}
Evidently, $\Theta_{\rm classical}$ is conjugate to the harmonic-oscillator
Hamiltonian in the sense of classical action-angle variables.

One can also interpret (\ref{e8}) at the quantum level by using the simple
identity \cite{S5,S6}
\begin{equation}
T_{-n,n}=\frac{1}{2}\left(q\frac{1}{p}\right)^n+\frac{1}{2}\left(\frac{1}{q}p
\right)^n.
\label{e11}
\end{equation}
This formula allows us to sum the series in (\ref{e8}) at the operator level
(where the ordering of operators is crucial):
\begin{eqnarray}
\Theta_{\rm min}&=&\sum_{k=0}^\infty\frac{(-1)^{k+1}}{4k+2}\left(p^{2k+1}
q^{-2k-1}+q^{-2k-1}p^{2k+1}\right)\nonumber\\
&=&\frac{1}{2}{\rm arctan}\left(p\frac{1}{q}\right)+\frac{1}{2}
{\rm arctan}\left(\frac{1}{q}p\right).
\label{e12}
\end{eqnarray}
It is important to observe that the formal summation in (\ref{e12}) produces a
{\it bounded} operator even though the individual terms in the summation are
unbounded operators. (The same effect can be seen in the expansion $e^{iq}=
\sum_{n=0}^\infty i^nq^n/n!$.) The fact that infinite sums of ill-behaved
objects can be well behaved will be used throughout this paper.

Observe that if the Hamiltonian $H$ is symmetric under parity reflection $[H,
\cP]=0$, then as a consequence of (\ref{e1}) the time operator $\Theta$ will
also be symmetric under parity reflection. Also, if $H$ is symmetric under time
reversal $[H,\cT]=0$, then $\Theta$ will be {\it anti}symmetric under $\cT$
because the time-reversal operator changes the sign of $i$. (Note that if $H$ is
non-Hermitian but $\cP\cT$ symmetric, then $\Theta$ will be $\cP\cT$ {\it
anti}symmetric \cite{S7}.) The Hamiltonian for the quantum harmonic oscillator
is both $\cP$ and $\cT$ symmetric because under parity reflection $p\to-p$ and
$q\to-q$ and under time reversal $p\to-p$, $q\to q$, and $i\to-i$. Thus, the
minimal solution in (\ref{e8}) and (\ref{e12}) exhibits the expected $\cP$
symmetry and $\cT$ antisymmetry.

The question addressed in this paper is similar to that addressed in
Ref.~\cite{S5,S6} in which the uniqueness of the $\cC$ operator in $\cP\cT$
quantum theory was examined \cite{S7}. Because the time operator is not unique,
our objective here is to determine whether there is a solution to (\ref{e1})
that is optimal in some sense. Our approach will be to calculate matrix elements
of the time operator $\Theta$ and to ask whether there is a choice of
homogeneous solution $\Theta_{\rm hom}$ which, when added to $\Theta_{\rm min}$,
gives particularly simple results.

In Sec.~\ref{s2} we construct series representations for the solutions to the
homogeneous commutator equation (\ref{e9}). Then in Sec.~\ref{s3} we show how to
use zeta-function regulation to express matrix elements of the singular
operators in these series representations as polynomials. Next, in Sec.~\ref{s4}
we discuss the detailed mathematical structure of the polynomials obtained in
Sec.~\ref{s3}. Finally, in Sec.~\ref{s5} we give some concluding remarks. In the
Appendix we explain the details of the zeta-function regulation techniques used
in Sec.~\ref{s3}.

\section{Homogeneous Solutions to Eq.~(\ref{e1}) for the Harmonic Oscillator}
\label{s2}

In this section we show how to obtain homogeneous (nonminimal) solutions to the
commutator equation (\ref{e1}) for the harmonic-oscillator Hamiltonian 
(\ref{e3}). We will seek solutions of the general form (\ref{e4}). These
solutions can be expressed as a single (not a double) sum on the
index $k$ and can be labeled by an integer parameter $\gamma$.

\subsection{Homogeneous solutions odd in $p$ and odd in $q$}
\label{ss2a}

In this paper we are only interested in homogeneous solutions that are odd in
$p$ and odd in $q$. The fact that $H$ in (\ref{e3}) is symmetric under space
reflection implies that $\Theta$ cannot be odd in $p$ and even in $q$, or even
in $p$ and odd in $q$. Furthermore, since $H$ is symmetric under time reversal,
$\Theta$ must be either real and odd in $p$ and odd in $q$, or else imaginary
and even in $p$ and even in $q$. In the latter case, however, $\Theta$ will just
turn out to be a function of $H$, and therefore the matrix elements of $\Theta$
in the harmonic-oscillator-eigenfunction basis will be diagonal. The point of
this paper is to use the homogeneous solutions to simplify the matrix elements
of the time operator, and therefore even-even solutions will not be of any
value. Thus, we focus here on odd-odd homogeneous solutions. Solutions having
this symmetry property can be written in the general form 
\begin{equation}
\Theta_{\rm hom}^{(\gamma)}=\sum_{k=-\infty}^\infty
a_k^{(\gamma)}T_{-2k+2\gamma-1,2k+1}.
\label{e13}
\end{equation}

Substituting (\ref{e13}) into (\ref{e9}) gives the recursion relation
\begin{equation}
(2k+1-2\gamma)a_{k}^{(\gamma)}+(2k+3)a_{k+1}^{(\gamma)}=0
\label{e14}
\end{equation}
for $-\infty<k<\infty$. Thus, 
\begin{equation}
a_k^{(\gamma)}=(-1)^ka_0^{(\gamma)}\frac{\sqrt{\pi}\,\Gamma(k-\gamma+1/2)}
{2\Gamma(1/2-\gamma)\Gamma(3/2+k)},
\label{e15}
\end{equation}
where $a_0^{(\gamma)}$ is an arbitrary constant for each value of $\gamma$.

\begin{figure}[h!]
\begin{center}
\includegraphics[scale=0.39, viewport=0 0 624 624]{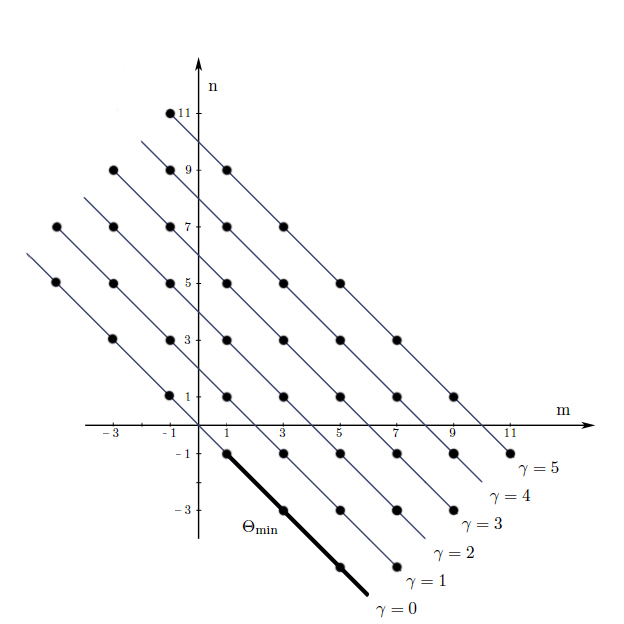} 
\end{center}
\caption{Schematic representation indicating the totally symmetric operators 
$T_{m,n}$ that contribute to the sum in (\ref{e8}) for $\Theta_{\rm min}$ and to
the sum in (\ref{e13}) for $\Theta_{\rm hom}^{(\gamma)}$. Each nonzero term in
these sums is indicated by a dot. For each value of $\gamma$ these dots lie on
diagonal lines. The dots associated with $\Theta_{\rm min}$ lie on a bold
diagonal line.}
\label{F1}
\end{figure}

The totally symmetric operators $T_{m,n}$ that contribute to $\Theta_{\rm hom}^{
(\gamma)}$ are shown in Fig.~\ref{F1} as dots. For each $\gamma$ these dots lie
on a diagonal line. The dots that contribute to $\Theta_{\rm min}^{(\gamma)}$
lie on a bold diagonal line.

To evaluate the sum in (\ref{e13}) for positive integers $\gamma$, we first use
the symmetry property $a_k^{(\gamma)}=a_{-k-1+\gamma}^{(\gamma)}$ to split the
sum into two sums. Then, we use the anticommutator relations in (\ref{e5}) to
obtain
\begin{widetext}
\begin{eqnarray}
\Theta_{\rm hom}^{(\gamma)}&=&\frac{1}{4^\gamma}\sum_{k=0}^\infty a_k^{(\gamma)}
\left[\left\{q,\ldots\left\{q,\left\{q,T_{2k+1,-2k-1}\right\}\right\}\ldots
\right\}_{(2\gamma)~{\rm times}}\right.\nonumber\\
&&\quad+\left.\left\{p,\ldots\left\{p,\left\{p,T_{-2k-1,2k+1}\right\}\right\}
\ldots\right\}_{(2\gamma)~{\rm times}}\right]+\sum_{k=0}^{\gamma-1}
a_k^{(\gamma)}T_{2k+1,-2k-1+2\gamma}.
\label{e16}
\end{eqnarray}
Finally, we use the identity (\ref{e11}) to simplify the totally symmetric
operators of the form $T_{-m,m}$. (This identity is valid for both positive- and
negative-integer $m$.) This gives the result
\begin{eqnarray}
\Theta_{\rm hom}^{(\gamma)}&=&\frac{1}{2^{2\gamma+1}}\sum_{k=0}^\infty a_k^{(
\gamma)}\left[\left\{q,\ldots\left\{q,\left\{q,\left(p\frac{1}{q}\right)^{2k+1}
+\left(\frac{1}{q}p\right)^{2k+1}\right\}\right\}\ldots\right\}
_{(2\gamma)~{\rm times}}\right.\nonumber\\
&&\quad+\left.\left\{p,\ldots\left\{p,\left\{p,\left(\frac{1}{p}q\right)^{2k+1}+
\left(q\frac{1}{p}\right)^{2k+1}\right\}\right\}\dots\right\}_{(2\gamma)~{\rm
times}}\right]+\sum_{k=0}^{\gamma-1}a_k^{(\gamma)}T_{2k+1,-2k-1+2\gamma},
\label{e17}
\end{eqnarray}
\end{widetext}
which we emphasize is only valid for positive-integer $\gamma$. It is possible
to sum each of the four infinite series in (\ref{e17}) in terms of a
hypergeometric function
\begin{equation}
\sum_{k=0}^\infty x^{2k}a_k^{(\gamma)}= a_0^{(\gamma)}{}_2{\rm F}_1\left(
\frac{1}{2}-\gamma,1,\frac{3}{2},-x^2\right),
\label{e18}
\end{equation}
but there is no simple way to sum the finite series in (\ref{e17}).

The coefficients in (\ref{e15}) are also valid for negative-integer $\gamma$,
but now because $\gamma$ is negative, we use the commutator relations in
(\ref{e5}) (instead of the anticommutator relations) to obtain the general
formula
\begin{eqnarray}
T_{-m-2\gamma,m} &=& \frac{(-1)^\gamma(m-1)!}{(m+2\gamma-1)!}\nonumber\\
&& \!\!\!\!\!\!\!\!\!\!\!\!\!\!\!\! \times\left[\ldots\left[\left[T_{-m,m},q
\right],q\right]\ldots,q\right]_{(2\gamma)~{\rm times}},
\label{e19}
\end{eqnarray}
and then we proceed as above.

It is easy to see that $\Theta_{\rm hom}^{(\gamma)}$ is {\it not} a function of
$H$ only. This is because the operator $\Theta_{\rm hom}^{(\gamma)}$ changes
sign under time reversal $\cT$ while $H$ does not change sign. Furthermore,
if $\Theta_{\rm hom}^{(\gamma)}$ were a function of $H$ only, then its matrix
elements in the harmonic-oscillator-eigenstate basis would be diagonal. We will
see in Sec.~\ref{s3} that the matrix elements are not diagonal.

\subsection{Homogeneous solutions odd in $p$ and even in $q$}
\label{ss2b}

In this paper we are only interested in homogeneous solutions that are odd in
both $p$ and $q$, but we point out that there are also solutions to (\ref{e9})
having other symmetry properties. For example, solutions that are odd in $p$ and
even in $q$ have the general form
\begin{equation}
\Theta_{\rm hom}^{(\gamma)}=\sum_k a_k^{(\gamma)}T_{2\gamma+1-2k,2k},
\label{e20}
\end{equation}
where the parameter $\gamma=0,\pm1,\pm2,\ldots$~.

Substituting (\ref{e20}) into (\ref{e9}), we obtain the following two-term
recursion relation for the coefficients $a_k^{(\gamma)}$:
\begin{equation}
a_{k+1}^{(\gamma)}(k+1)-(\gamma-k+1/2)a_k^{(\gamma)}=0\quad(k=0,1,2,\ldots).
\label{e21}
\end{equation}
Unlike the recursion relation (\ref{e14}), this recursion relation is
self-terminating; that is, if we choose $a_{-1}^{(\gamma)}=0$, then $a_k^{(
\gamma)}=0$ for $k<0$, $a_0^{(\gamma)}$ is arbitrary, and $a_k^{(\gamma)}\neq0$
for $k>0$. In terms of $a_0^{(\gamma)}$ the solution to this recursion relation
is
\begin{equation}
a_k^{(\gamma)}=a_0^{(\gamma)}(-1)^k\frac{\Gamma(k-\gamma-1/2)}{k!\Gamma(-\gamma-
1/2)}\quad(k=0,1,2,\ldots).
\label{e22}
\end{equation}

The series (\ref{e20}) with coefficients (\ref{e22}) can be summed as a
binomial expansion:
\begin{equation}
\sum_{k=0}^\infty a_k^{(\gamma)}x^{2k}=a_0^{(\gamma)}\left(1+x^2\right)^{\gamma+1/2}.
\label{e23}
\end{equation}
Thus, for $\gamma\geq0$ the odd-even one-parameter family of solutions to the
homogeneous equation (\ref{e9}) is
\begin{widetext}
\begin{equation}
\Theta_{\rm hom}^{(\gamma)}=\frac{a_0^{(\gamma)}}{2^{2\gamma+2}}\left\{\ldots
\left\{\left\{\left(1+q\frac{1}{p}q\frac{1}{p}\right)^{\gamma+1/2}
+\left(1+\frac{1}{p}q\frac{1}{p}q\right)^{\gamma+1/2}
,p\right\},p\right\}\dots,p\right\}_{(2\gamma+1)~{\rm times}}.
\label{e24}
\end{equation}
For example, for $\gamma=0$
\begin{equation}
\Theta_{\rm hom}^{(0)}=\frac{1}{4}a_0^{(0)}\left(\sqrt{1+q\frac{1}{p}q
\frac{1}{p}}\,p+p\,\sqrt{1+q\frac{1}{p}q\frac{1}{p}}+\sqrt{1+\frac{1}{p}q
\frac{1}{p}q}\,p+p\,\sqrt{1+\frac{1}{p}q\frac{1}{p}q}\right).
\label{e25}
\end{equation}
\end{widetext}
For $\gamma<0$ the result is similar, but it involves commutators. For $\gamma=
-1$, we get
\begin{equation}
\Theta_{\rm hom}^{(-1)}=\frac{a_0^{(-1)}}{2i}\left[p,{\rm arcsinh}\left(q
\frac{1}{p}\right)+{\rm arcsinh}\left(\frac{1}{p}q\right)\right].
\label{e26} 
\end{equation}

We can see again from (\ref{e24})--(\ref{e26}) that, as we stated in
Sec.~\ref{s1}, while $\Theta_{\rm hom}^{(\gamma)}$ commutes with the Hamiltonian
$H$ in (\ref{e3}), it is not a function of $H$. We verify this by observing that
$\Theta_{\rm hom}^{(\gamma)}$ is odd under parity reflection $\cP$ while $H$ is
even. Moreover, in the classical limit, where $p$ and $q$ commute, (\ref{e25})
simplifies to 
\begin{equation}
\Theta_{\rm hom}^{(0)}=a_0^{(0)}{\rm sgn}(p)\sqrt{2H}.
\label{e27}
\end{equation}

The solutions in (\ref{e24})--(\ref{e27}) are solutions to the commutator
equation (\ref{e9}), but they are physically unacceptable because they violate
time-reversal symmetry. Thus, they must be rejected. However, the
homogeneous solutions in (\ref{e13}), whose numerical coefficients are given in
(\ref{e15}), are physically acceptable because they exhibit the correct behavior
under time and space reflection. This raises the question of which solution for
the time operator $\Theta$ is optimal and in what sense it is optimal. We
address this issue in the next section.

\section{Matrix elements of the time operator}
\label{s3}

We showed in Sec.~\ref{s2} that there are many possible choices for the time
operator $\Theta$ because there are an infinite number of one-parameter families
of homogeneous solutions to (\ref{e9}). One criterion for deciding on the
optimal choice for the time operator (if there is an optimal choice) is
mathematical simplicity. To investigate this criterion, in this section we
calculate the matrix elements of the time operator in a basis consisting of
harmonic-oscillator eigenfunctions. We begin by calculating the matrix elements
of the minimal solution (\ref{e8}). Then, we ask whether these matrix elements
simplify when combined with homogeneous solutions of the type in (\ref{e13}).

In coordinate space the $n$th normalized eigenfunction of the harmonic
oscillator is
\begin{equation}
\psi_n(q)=\frac{1}{\pi^{1/4}\sqrt{2^n n!}}e^{-q^2/2}H_n(q).
\label{e28}
\end{equation}
If we try to calculate the matrix elements of $\Theta_{\rm min}$ or $\Theta_{\rm
hom}$ in this basis, we run into difficulties because each term in the series
representation for these operators is singular (because powers of $1/p$ or $1/q$
appear.) However, in this section we propose a method for circumventing these
difficulties and thus for obtaining a finite result for the matrix elements.
Specifically, we argue that while individual terms in the series representation
for the time operator are singular, the sum of the series is a bounded operator
[see, for example, (\ref{e12})]. We then show that the $(m,n)$ matrix element of
a given singular operator in the sum can be made finite depending on the
evenness or oddness of $m$ and $n$ and also depending on whether the singular
operator is taken to act to the left or to the right.

\subsection{Matrix representation of the minimal solution}
\label{ss3a}

To calculate the $(m,n)$ matrix element of $\Theta_{\rm min}$, we must
calculate the matrix element of each individual term $T_{2k+1,-2k-1}$ in the
series (\ref{e8}). The parity of this operator is even, so only the nonzero
matrix elements correspond to values of $m$ and $n$ that are both odd or both
even. Referring to (\ref{e11}), we can see that in coordinate space the $(m,n)$
matrix element of $T_{2k+1,-2k-1}$ has the form
\begin{eqnarray}
&&\frac{1}{2}(-i)^{2k+1}\int_{-\infty}^\infty dq\,\psi_m(q)\left(\frac{1}{q}
\partial_q\right)^{2k+1}\psi_n(q)\nonumber\\
&&+\frac{1}{2}(-i)^{2k+1}\int_{-\infty}^\infty\psi_m(q)\left(\partial_q
\frac{1}{q}\right)^{2k+1}\psi_n(q).
\label{e29}
\end{eqnarray}
Note that the first integral exists if $n$ is even and the second integral
exists if $n$ is odd. When $n$ is even, we reinterpret the differential operator
in the second integral as operating to the left, and when $n$ is odd we
reinterpret the differential operator in the first integral as operating to the
left. (Note that when the operator acts to the left, there is an additional
minus sign because $2k+1$ is odd.)

\begin{figure}[t!]
\begin{center}
\vspace{.3cm}
\includegraphics[scale=0.55, viewport=0 0 441 354]{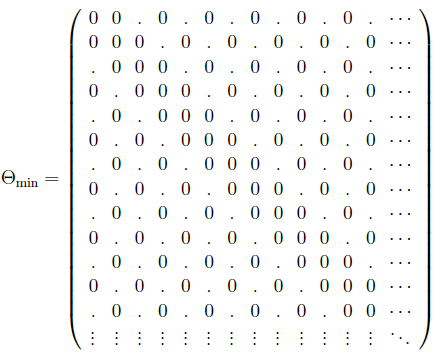}
\end{center}
\caption{Schematic view of the matrix elements of the minimal time operator
$\Theta_{\rm min}$. The nonvanishing matrix elements are shown as dots.}
\label{F2}
\end{figure}

The only nonzero matrix elements of $\Theta_{\rm min}$ are
\begin{eqnarray}
&& \langle 2n-2j|\Theta_{\rm min}|2n\rangle\nonumber\\
&&\quad =
-\frac{i}{2}\sum_{k=0}^\infty\frac{1}{2k+1}\langle 2n-2j|\left(\frac{1}{q}
\partial_q \right)^{2k+1}|2n\rangle,\nonumber \\
&& \langle 2n-2j-1|\Theta_{\rm min}|2n-1\rangle\nonumber\\
&& \quad =
-\frac{i}{2}\sum_{k=0}^\infty\frac{1}{2k+1}\langle 2n-2j-1|\left(\partial_q
\frac{1}{q}\right)^{2k+1}|2n-1\rangle,\nonumber \\
&& \langle 2n|\Theta_{\rm min}|2n-2j\rangle =
-\langle 2n-2j|\Theta_{\rm min}|2n\rangle,\nonumber \\
&& \langle 2n-1|\Theta_{\rm min}|2n-2j-1\rangle\nonumber\\
&& \quad =
-\langle 2n-2j-1|\Theta_{\rm min}| 2n-1\rangle,
\label{e30}
\end{eqnarray}
where $j=1,2,3\ldots$ and $n\geq j$. Note that the diagonal matrix elements
vanish: $\langle m|\Theta_{\rm min}|m\rangle=0$. In Fig.~\ref{F2} the nonzero
matrix elements are shown as dots and the vanishing matrix elements are shown as
zeros.

The calculation of the matrix elements in (\ref{e30}) is lengthy, and we have
relegated the discussion to the Appendix. Here, we merely report the results of
the calculation. All the even-even and odd-odd matrix elements of the minimal
solution (\ref{e8}) can be expressed compactly as
\begin{eqnarray}
\langle 2n-2j|\Theta_{\rm min}|2n\rangle &=& i (-1)^j 2^{j/2-1}(j-1)!\nonumber\\
&&\!\!\!\!\!\!\!\!\!\!\!\!\!\!\!\!\!\!\!\!\!\!\!\!\!
\times\sqrt{\frac{n!(2n-2j-1)!!}{(n-j)!(2n-1)!!} } F(j),\nonumber\\
\langle 2n+1-2j|\Theta_{\rm min}|2n+1\rangle &=& i
(-1)^{j+1}2^{j/2-1}(j-1)!\nonumber\\
&&\!\!\!\!\!\!\!\!\!\!\!\!\!\!\!\!\!\!\!\!\!\!\!\!\!
\times\sqrt{\frac{n!(2n-2j+1)!!}{(n-j)!(2n+1)!!} }F(j),
\label{e31}
\end{eqnarray}
where
\begin{equation}
F(j)=\sum_{\rho=1}^j\frac{2^{\rho}}{\Gamma(\rho)\Gamma(1-\rho+j)}Z(\rho),
\label{e32}
\end{equation}
\begin{equation}
Z(\rho)=\frac{\delta_{\rho,1}}{2}-\sum_{i=1}^\rho\frac{S_{\rho,i}^1}{\rho!}
\sum_{r=1}^{\lfloor i/2+1\rfloor}\binom{i-1}{2r-1}\ 2^{2r-2}\frac{\cB_{2r}}{r},
\label{e33}
\end{equation}
$\cB_{2r}$ are Bernoulli numbers, and $S_{\rho,i}^1$ are Stirling numbers of
the first kind. Table \ref{t1} gives the first eight values of the functions
$Z(\rho)$ and $F(j)$.

\begin{table}[t!]
\begin{tabular}{|c|c|}
\hline $j$ & $F(j)$\tabularnewline
\hline \hline 1 & 1\tabularnewline
\hline 2 & 2/3\tabularnewline
\hline 3 & 5/18\tabularnewline
\hline 4 & 11/135\tabularnewline
\hline 5 & 101/5400\tabularnewline
\hline 6 & 593/170100\tabularnewline
\hline 7 & 2623/4762800\tabularnewline
\hline 8 & 383/5103000\tabularnewline
\hline
\end{tabular}
\begin{tabular}{|c|c|}
\hline
$\rho$ & $Z(\rho)$\tabularnewline
\hline \hline 1 & 1/2\tabularnewline
\hline 2 & -1/12\tabularnewline
\hline 3 & 1/36\tabularnewline
\hline 4 & -1/90\tabularnewline
\hline 5 & 1/200\tabularnewline
\hline 6 & -113/45360\tabularnewline
\hline 7 & 29/21168\tabularnewline
\hline 8 & -47/56700\tabularnewline
\hline
\end{tabular}
\caption{First eight values of the functions $F(j)$ in (\ref{e32}) and
$Z(\rho)$ in (\ref{e33}).}
\label{t1}
\end{table}

The explicit values of some matrix elements are then
\begin{eqnarray}
\langle2n-1|\Theta_{\rm min}|2n+1\rangle
&=&\frac{i}{2}\sqrt{\frac{2n}{(2n+1)}},\nonumber\\
\langle2n-2|\Theta_{\rm min}|2n\rangle
&=&-\frac{i}{2}\sqrt{\frac{2n}{(2n-1)}},\nonumber\\
\langle2n-3|\Theta_{\rm min}|2n+1\rangle
&=&-\frac{2i}{3}\sqrt{\frac{n(n-1)}{(2n+1)(2n-1)}},\nonumber\\
\langle2n-4|\Theta_{\rm min}|2n\rangle
&=&\frac{2i}{3}\sqrt{\frac{n(n-1)}{(2n-1)(2n-3)}},\nonumber
\end{eqnarray}

\begin{eqnarray}
\langle2n-5|\Theta_{\rm min}|2n+1\rangle &=& \nonumber\\ &&
\!\!\!\!\!\!\!\!\!\!\!\!\!\!\!\!\!\!\!\!\!\!\!\!\!\!\!\!\!\!\!\!\!\!\!\!\!\!\!\!
\frac{5i}{9}\sqrt{\frac{2n(n-1)(n-2)}{(2n+1)(2n-1)(2n-3)}},\nonumber \\
\langle2n-6|\Theta_{\rm min}|2n\rangle &=& \nonumber\\ &&
\!\!\!\!\!\!\!\!\!\!\!\!\!\!\!\!\!\!\!\!\!\!\!\!\!\!\!\!\!\!\!\!\!\!\!\!\!\!\!\!
-\frac{5i}{9}\sqrt{\frac{2n(n-1)(n-2)}{(2n-1)(2n-3)(2n-5)}},\nonumber \\
\langle 2n-7|\Theta_{\rm min}|2n+1\rangle &=& \nonumber\\ &&
\!\!\!\!\!\!\!\!\!\!\!\!\!\!\!\!\!\!\!\!\!\!\!\!\!\!\!\!\!\!\!\!\!\!\!\!\!\!\!\!
-\frac{44i}{45}\sqrt{\frac{n(n-1)(n-2)(n-3)}{(2n+1)(2n-1)(2n-3)(2n-5)}},
\nonumber \\
\langle 2n-8|\Theta_{\rm min}|2n\rangle &=& \nonumber\\ &&
\!\!\!\!\!\!\!\!\!\!\!\!\!\!\!\!\!\!\!\!\!\!\!\!\!\!\!\!\!\!\!\!\!\!\!\!\!\!\!\!
\frac{44i}{45}\sqrt{\frac{n(n-1)(n-2)(n-3)}{(2n-1)(2n-3)(2n-5)(2n-7)}}.\nonumber
\end{eqnarray}

\subsection{Matrix representations of the homogeneous solutions}
\label{ss3b}

The question is whether we can simplify the result in (\ref{e31}) by including
the homogeneous solutions in (\ref{e13}). Let us first examine the contribution
from the homogeneous solution corresponding to $\gamma=0$. We find that the
nonzero matrix elements for $\Theta_{\rm hom}^{(0)}$ lie on {\it alternating}
diagonals in Fig.~\ref{F2}; that is, the matrix elements are nonzero on the
first dotted diagonals immediately above and immediately below the main
diagonal. On the next dotted diagonals immediately above and below these dotted
diagonals, the matrix elements vanish; on the next dotted diagonals immediately
above and below these dotted diagonals, the matrix elements do not vanish; and
so on. Specifically, we find that 
\begin{equation}
\langle n-4j+2 |\Theta_{\rm hom}^{(0)}|n\rangle=-2 a_0^{(0)}\langle n-4j+2
|\Theta_{\rm min}|n\rangle
\label{e34}
\end{equation}
for $j=1,2,3,\ldots$ and $n\geq4j-2$. On the diagonals on which the matrix
elements of $\Theta_{\rm hom}^{(0)}$ do not vanish, each matrix element is a
multiple of the matrix elements of $\Theta_{\rm min}$. Hence, we can choose
$a_0^{(0)}=1/2$ in (\ref{e15}) and cancel half of the nonzero matrix elements of
$\Theta_{\rm min}$. Thus, including this homogeneous solution causes a dramatic
simplification of the matrix elements of the time operator!

Next, we consider the matrix elements of $\Theta_{\rm hom}^{(\gamma)}$ for
$\gamma\geq1$. Here are some useful properties of the matrix representation of
the operators $T_{m,n}$:
\begin{eqnarray}
\langle n-4j |T_{m,k}|n\rangle &=& -\langle n-4j |T_{k,m}|n\rangle,
\nonumber\\
\langle n-4j |T_{k,k}|n\rangle &=& 0,
\label{e35}
\end{eqnarray}
for $n \geq 4j$ and  for $k,m \geq 1$. For fixed $\gamma$, only the columns with
$j\leq\lfloor\frac{\gamma+1}{2}\rfloor$ are not empty. This means that the
nonzero matrix elements of the operators $\Theta_{\rm hom}^{(\gamma)}$ for
$\gamma=1$ and $\gamma=2$ lie on just two diagonals above and two diagonals
below the main diagonal; the nonzero matrix elements for $\gamma=3$ and $\gamma=
4$ lie on two and six diagonals above and below the main diagonal; the nonzero
matrix elements for $\gamma=5$ and $\gamma=
6$ lie on two, six, and ten diagonals above and below the main diagonal; and so
on. (The diagonals containing nonzero matrix elements are indicated in
Fig.~\ref{F3} by dots.) To summarize,
\begin{widetext}
\begin{equation}
\langle n-4j+2 |\Theta_{\rm hom}^{(2j+\sigma-1)}|n\rangle=\sum_{k=0}^{2j+\sigma
-2} a_k^{(2j+\sigma-1)}\langle n-4j+2 |T_{2k+1,2(2j+\sigma-k-1)-1}|n\rangle,
\label{e36}
\end{equation}
where we have substituted $\gamma=2j+\sigma-1$ with $\sigma=0,1,2,\ldots$. The
coefficients $a_k^{(\gamma)}$ are given in (\ref{e15}).

The matrix elements of the operators $T_{m,n}$ are associated with interesting
classes of orthogonal polynomials. In particular, we are now interested in the
case $m,n$ odd. The $j$th class of orthogonal polynomials $\left\{\cP^{(j)}_{k}
(n)\right\}_{k=0,1\ldots}$ is defined by the equation
\begin{equation}
\langle n-2j|T_{2k+1,2\ell+1}|n\rangle=-i\sqrt{\frac{n!}{(n-2j)!}}\,\cC_
{\ell,k}^{(j)}\,\cP^{(j)}_{\ell +k-j+1}(n)\quad(k,\ell\geq 0,\,j\geq1,\,
n\geq 2j)
\label{e37}
\end{equation}
where
\begin{equation}
\cC_{\ell, k}^{(j)}=\frac{(-1)^kj(\ell+k)!(2k+2)!(2\ell+2)!}{2^{\ell
+k+2}(\ell+1)!(\ell +k+j+1)!(\ell+k-j+1)!(k+1)!}.
\label{e38}
\end{equation}
The polynomials $\cP^{(j)}_{k}(n)$ ($j=1,\,2,\,\ldots$) have degree $k$ and
argument $n$ and are discussed in detail in the next section. Substituting
(\ref{e37}) with $j\to 2j-1$ and $\ell=2j+\sigma-k-2$ into (\ref{e36}), we
obtain the following expression for the nonvanishing diagonals of elements in
the matrix representation of the homogeneous solutions $\Theta_{\rm hom}^{(
\gamma)}$:
\begin{equation}
\langle n-4j+2|\Theta_{\rm hom}^{(2j-1+\sigma)}|n\rangle=-i a_0^{(\sigma+2j-1)}
\frac{(2j-1)(2\sigma+4j-2)!}{2^{2j+\sigma-1}(\sigma+4j-2)!\sigma!}\sqrt{
\frac{n!}{(n-4j+2)!}}\cP^{(2j-1)}_\sigma(n),
\label{e39}
\end{equation}
where $j=1,2,3,\ldots$, $n\geq 4j-2$, and $\sigma=0,1,2,\ldots$.

\begin{figure}[t!]
\begin{center}
\includegraphics[scale=0.30]{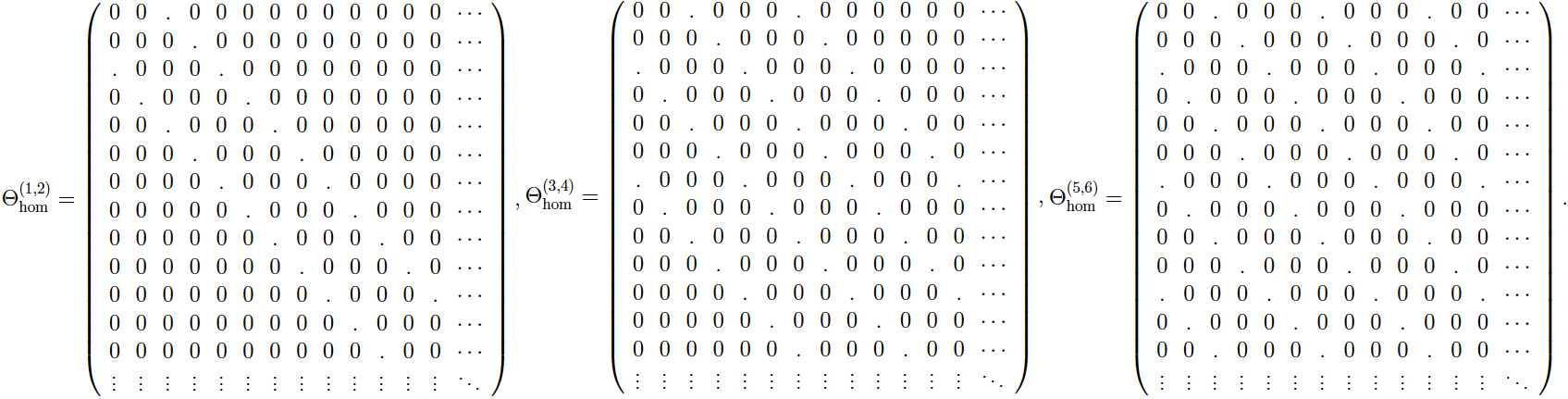}
\end{center}
\caption{Schematic representation of the matrix elements for homogeneous
solutions $\Theta_{\rm hom}^{\gamma}$ for $\gamma=1,2,\ldots,6$. The dots
indicate the nonzero matrix elements. For $\gamma=1$ and $\gamma=2$ there are
only two diagonals where the matrix elements are nonzero. When $\gamma=3,4$
there are four diagonals of nonzero matrix elements, when $\gamma=5,6$ there are
six diagonals, and so on.}
\label{F3}
\end{figure}
\end{widetext}

\section{Polynomial Representations of the Homogeneous solutions}
\label{s4}

In this section we describe the properties of the polynomials in (\ref{e37}).
The polynomials $\cP^{(j)}_k(x)$ are orthogonal and satisfy the
three-term recursion relation
\begin{eqnarray}
\cP^{(j)}_{k+1}(x)&=&\frac{1}{2}(2x-2j+1)\cP^{(j)}_k(x)
\nonumber\\
&&+\frac{k}{4}(k+2j)\cP^{(j)}_{k-1}(x),
\label{e40}
\end{eqnarray}
where $\cP^{(j)}_0(x)=1$. The generating function
\begin{equation}
G^{(j)}(t)=\sum_{k=0}^\infty \frac{t^k}{k!}\cP^{(j)}_k(x)
\label{e41}
\end{equation}
for these polynomials is
\begin{equation}
G^{(j)}(t)=2^{2j+1}\frac{(2+t)^{x-2j}}{(2-t)^{x+1}}.
\label{e42}
\end{equation}
From this generating function we identify these polynomials as Meixner
polynomials of the first kind \cite{S9,S10}, whose definition and properties are
elucidated in the next subsection. In this paper $j\in\mathbb{N}^+$, but if we
set $j=0$, we obtain the Hahn polynomials that were studied in Ref.~\cite{S11}.

The polynomials $\cP^{(j)}_k(x)$ do not have parity. However, if we shift the
argument by $x=iy+(2j-1)/2$ and define $\cQ^{(j)}_k(y)=(-i)^k\cP^{(j)}_k[iy+(2j-
1)/2]$, then the new polynomials {\it do} have parity. Below we give the first
six polynomials $\cQ^{(j)}_k(y)$ for $j=1$ through $j=5$:
\begin{eqnarray}
\cQ^{(1)}_0(y) &=& 1,\qquad\cQ^{(1)}_1(y)\,\,=\,\,y,\nonumber\\
\cQ^{(1)}_2(y) &=& y^2-\frac{3}{4},\qquad\cQ^{(1)}_3(y)\,\,=\,\,
y^3-\frac{11}{4}y,\nonumber\\
\cQ^{(1)}_4(y) &=& y^4-\frac{13}{2}y^2+\frac{45}{16},\nonumber\\
\cQ^{(1)}_5(y) &=& x^5-\frac{25}{2}y^3+\frac{309}{16},
\label{e43}
\end{eqnarray}

\begin{eqnarray}
\cQ^{(2)}_0(y) &=& 1,\qquad\cQ^{(2)}_1(y)\,\,=\,\,y,\nonumber\\
\cQ^{(2)}_{2}(y) &=&y^2-\frac{5}{4},\qquad\cQ^{(2)}_{3}(y)
\,\,=\,\,y^3-\frac{17}{4}y,\nonumber\\
\cQ^{(2)}_{4}(y) &=&y^4-\frac{19}{2}y^2+\frac{105}{16},\nonumber\\
\cQ^{(2)}_{5}(y) &=&y^5-\frac{35}{2}y^3+\frac{649}{16}y,
\label{e44}
\end{eqnarray}

\begin{eqnarray}
\cQ^{(3)}_0(y) &=& 1,\qquad\cQ^{(3)}_1(y)\,\,=\,\,y,\nonumber\\
\cQ^{(3)}_{2}(y) &=&y^2-\frac{7}{4},\qquad\cQ^{(3)}_{3}(y)
\,\,=\,\,y^3-\frac{23}{4}y,\nonumber\\
\cQ^{(3)}_{4}(y) &=&y^4-\frac{25}{2}y^2+\frac{189}{16},\nonumber\\
\cQ^{(3)}_{5}(y) &=&y^5-\frac{45}{2}y^3+\frac{1109}{16}y,
\label{e45}
\end{eqnarray}

\begin{eqnarray}
\cQ^{(4)}_0(y) &=& 1,\qquad\cQ^{(4)}_1(y)\,\,=\,\,y,\nonumber\\
\cQ^{(4)}_{2}(y) &=& y^2-\frac{9}{4},\qquad\cQ^{(4)}_{3}(y)
\,\,=\,\,y^3-\frac{29}{4}y,\nonumber\\
\cQ^{(4)}_{4}(y) &=& y^4-\frac{31}{2}y^2+\frac{297}{16},\nonumber\\
\cQ^{(4)}_{5}(y) &=& y^5-\frac{55}{2}y^3+\frac{1689}{16}y,
\label{e46}
\end{eqnarray}

\begin{eqnarray}
\cQ^{(5)}_0(y) &=& 1,\qquad\cQ^{(5)}_1(y)\,\,=\,\,y,\nonumber\\
\cQ^{(5)}_{2}(y) &=& y^2-\frac{11}{4},\qquad\cQ^{(5)}_{3}(y)
\,\,=\,\,y^3-\frac{35}{4}y,\nonumber\\
\cQ^{(5)}_{4}(y) &=& y^4-\frac{37}{2}y^2+\frac{429}{16},\nonumber\\
\cQ^{(5)}_{5}(y) &=& y^5-\frac{65}{2}y^3+\frac{2389}{16}y,
\label{e47}
\end{eqnarray}

The polynomials $\cQ^{(j)}_k(y)$ are orthogonal on $(-\infty,\infty)$
with respect to an even, positive-weight function $w^{(j)}(y)$:
\begin{equation}
\int_{-\infty}^\infty dy\,w^{(j)}(y)\cQ^{(j)}_r(y)\cQ^{(j)}_s(y)=0
\label{e48}
\end{equation}
for $r\neq s$. We use the notation
\begin{equation}
\mu^{(j)}_k=\int_{-\infty}^\infty dy\,w^{(j)}(y)y^{2k},
\label{e49}
\end{equation}
to represent the moments of the weight functions $w^{(j)}(y)$. These moments are
closely related to the Euler numbers $E_n$. In particular, for the case $j=0$
(see Ref.~\cite{S6})
\begin{equation}
\mu^{(0)}_k=|E_{2k}|4^{-k}.
\label{e50}
\end{equation}
In Table \ref{t2} we list the values of the moments for $j$ ranging from 1
through 6 and for $k$ ranging from 0 through 6.

The moments defined in (\ref{e49}) and displayed in Table \ref{t2} have some
remarkable mathematical properties. To list these properties we first express
these moments as polynomials ${\cal R}_k(j)=4^k\mu^{(j)}_k/(2j+1)$ of degree $k$
in the argument $j$. The first six of these polynomials are
\begin{eqnarray}
{\cal R}_1(j)&=&1,\nonumber\\
{\cal R}_2(j)&=&6j+5,\nonumber\\
{\cal R}_3(j)&=&60j^2+120j+61,\nonumber\\
{\cal R}_4(j)&=&840j^3+2940j^2+3486j+1385,\nonumber\\
{\cal R}_5(j)&=&15120j^4+80640j^3+163800j^2\nonumber\\
&&\!\!\!\!\!\!\!\!\!\!\!\!\!\!\!\!\!\! +148800j+50512,\nonumber\\
{\cal R}_6(j)&=&332640j^5+2494800j^4+7595280j^3\nonumber\\
&&\!\!\!\!\!\!\!\!\!\!\!\!\!\!\!\!\!\!+ 11641080j^2+8910726j+2702765.
\label{e51}
\end{eqnarray}

We then observe that these new polynomials resemble the Bernoulli polynomials
and the Euler polynomials in that we recover the Bernoulli numbers $B_{2k}$ and
the Euler numbers $E_{2k}$ when we evaluate the polynomials at special values of
$j$. As we saw in (\ref{e50}), we obtain the Euler number $|E_{2k}|$ when we
evaluate ${\cal R}_k(j)$ at $j=0$:
\begin{equation}
{\cal R}_k(0)=|E_{2k}|.
\label{e52}
\end{equation}
We obtain the Bernoulli numbers when we evaluate ${\cal R}_k(j)$ at $j=-1/2$:
\begin{equation}
{\cal R}_k(-1/2)=\frac{|B_{2k}|}{2k}\left(4^k-1\right)4^k.
\label{e53}
\end{equation}
There are many other values of $j$ for which the polynomials ${\cal R}_k(j)$
give simple results:
\begin{eqnarray}
{\cal R}_k(-3/2)&=&(-4)^{k-1},\nonumber\\
{\cal R}_k(-1)&=&(-1)^{k-1},\nonumber\\
{\cal R}_k(1/2)&=&\frac{|B_{2k+2}|}{k+1}\left(4^{k+1}-1\right)4^k,\nonumber\\
{\cal R}_k(-5/2)&=&(-1)^{k-1}2^{2k-3}\left(2^{2k-2}+1\right),\nonumber\\
{\cal R}_k(-2)&=&\frac{(-1)^{k+1}}{4}\left(3^{2k-1}+1\right).
\label{e54}
\end{eqnarray}

\begin{widetext}
\begin{table}[h!]
\begin{tabular}{|c|c|c|c|c|c|c|}
\hline
$k$ & $4^k\mu^{(1)}_k$ & $4^k\mu^{(2)}_k$ & $4^k\mu^{(3)}_k$ &
$4^k\mu^{(4)}_k$ & $4^k\mu^{(5)}_k$ & $4^k\mu^{(6)}_k$\tabularnewline
\hline \hline 0 & 1 & 1 & 1 & 1 & 1 & 1\tabularnewline
\hline 1 & 3 & 5 & 7 & 9 & 11 & 13\tabularnewline
\hline 2 & 33 & 85 & 161 & 261 & 385 & 533\tabularnewline
\hline 3 & 723 & 2705 & 6727 & 13509 & 23771 & 38233\tabularnewline
\hline 4 & 25953 & 134185 & 426881 & 1045161 & 2170465 & 4024553
\tabularnewline
\hline 5 & 1376643 & 9451805 & 37611847 & 110683809 & 268614731 &
570100453\tabularnewline
\hline 6 & 101031873 & 892060285 & 4355312801 & 15209937261 &
42750877345 & 103164046973\tabularnewline
\hline
\end{tabular}
\caption{Numerical values of the moments $\mu^{(j)}_k$ defined in (\ref{e49}).}
\label{t2}
\end{table}
\end{widetext}

\subsection{Connection between ${\cal P}_k^{(j)}(x)$ and
Meixner/Meixner-Pollaczek polynomials.}
\label{ss4a}

The polynomials ${\cal P}_k^{(j)}(x)$ defined in (\ref{e37}) satisfy the
three-term recursion (\ref{e40}) and are related to the Meixner polynomials:
\begin{equation}
{\cal P}_{k}^{(j)}(x)=2^{-k}{\cal M}_k(x-2j,\,2j+1,\,-1),
\label{e55}
\end{equation}
where the Meixner polynomials $\mathcal{M}_n(x,b,c)$ belong to the classical
orthogonal polynomials of a discrete variable and may be defined in terms of
hypergeometric functions:
\begin{equation}
\mathcal{M}_n(x,b,c)=\frac{\Gamma(b+k)}{\Gamma(b)}\phantom{}_2F_1\left(-k,-x;b;
1-c^{-1}\right).
\label{e56}
\end{equation}
For real values of the parameters $b>0$ and $0<c<1$ they satisfy a discrete
orthogonality relation with weight function
\begin{equation}
w(x)=\frac{\Gamma(b+x)c^x}{\Gamma(b)x!},
\label{e57}
\end{equation}
whose moments are 
\begin{equation}
\tilde{\mu}_k^{(b,c)}=\sum_{n=0}^{\infty}\frac{\Gamma(b+n)\,c^n n^k}{\Gamma(b)
\,n!}.
\label{e58}
\end{equation}
The problem with evaluating the moments associated with the polynomials
$\cP_k^{(j)}(x)$, that is, the moments of the Meixner polynomials with
$c=-1$, can be solved by performing a zeta-function regularization of the
divergent series in (\ref{e58}):
\begin{equation}
\tilde{\mu}_k^{(b,-1)}=\lim_{s\to 0}\sum_{n=1}^\infty
\frac{(-1)^n\Gamma(b+n)n^{k+s}}{\Gamma(b)n!}\quad(k\neq0).
\label{e59}
\end{equation}

The polynomials ${\cal Q}_k^{(j)}(x)$ are defined as
\begin{equation}
{\cal Q}_k^{(j)}(x)\equiv (-i)^k{\cal P}_k^{(j)}[ix+(2j-1)/2],
\label{e60}
\end{equation}
which are listed in (\ref{e43})-(\ref{e47}), satisfy the three-term recursion
relation ${\cal Q}_{k+1}^{(j)}(x)=x\,{\cal Q}_k^{(j)}(x)-\frac{k}{4}(k+2j)\,
{\cal Q}_{k-1}^{(j)}(x)$, and are connected with the Meixner-Pollaczek
polynomials $P_{k}^{\lambda}\left(x;\phi\right)$ \cite{S12} 
\begin{equation}
\cQ_k^{(j)}(x)=2^{-k}k!\,P_k^{(j+1/2)}(x;\,\pi/2).
\label{e61}
\end{equation}
The Meixner-Pollaczek polynomials are defined as
\begin{equation}
P_k(x,\phi)\equiv\frac{\Gamma(2\lambda+k)e^{ik\phi}}{\Gamma(2\lambda)\,k!}
{}_2F_1\left(-k,ix+\lambda;2\lambda;1-e^{-2i\phi}\right).
\label{e62}
\end{equation}
For $0<\phi<\pi$ and $\lambda>0$, $P_k(x,\phi)$ are continuous orthogonal
polynomials on the real line with respect to a continuous positive real measure
given by the weight function
\begin{equation}
w(x)=\frac{1}{2\pi}\left( 2 \sin\phi\right)^{2\lambda}|\Gamma(\lambda+ix)|^2
e^{x(2\phi-\pi)}.
\label{e63}
\end{equation}

The Meixner-Pollaczek polynomials $P^{\lambda}_k(x)$ are analytic continuations
of the Meixner polynomials $\mathcal{M}_n(x,b,c)$ in the parameter $c$
\cite{S13}. The connection is given by
\begin{equation}
P^{\lambda}_k(x)=e^{-ik\phi}\mathcal{M}_k(ix-\lambda,\,2\lambda,\,
e^{-2i\phi})/k!,
\label{e64}
\end{equation}
which is like the relation between $\cQ_k^{(j)}(x)$ and
$\cP_k^{(j)}(x)$.

\subsection{Evaluation of moments associated with ${\cal Q}_k^{(j)}(x)$}
\label{ss4b}

Equations (\ref{e61}) and (\ref{e64}) show the relationship between ${\cal
Q}_k^{(j)}(x)$ and ${\cal M}_{k}(x)$:
\begin{equation}
{\cal Q}_k^{(j)}(x)=(-i)^k 2^{-k}{\cal M}_k (ix-j-1/2,\,2j+1,\,-1).
\label{e65}
\end{equation}
Moments of the polynomials $\cQ_k^{(j)}(x)$ can be evaluated in two
ways. One way is to use (\ref{e64}) to express the moments $\mu_k^{(j)}$
associated with the polynomials ${\cal Q}_k ^{(j)}(x)$ in terms of the
regularized moments in (\ref{e59}) associated with the Meixner polynomials
\begin{eqnarray}
\mu_k^{(j)}&=&(-i)^k\sum_{\alpha=0}^k\binom{k}{\alpha}\,\left(j+1/2\right)^{
\alpha}\nonumber\\
&&\quad\times\sum_{n=0}^\infty\frac{(-1)^n\Gamma(2j+n+1)}{n!\,\Gamma(2j+1)}n^{k-
\alpha}.
\label{e66}
\end{eqnarray}

For $(2j+1)>0$ we can express the gamma function in (\ref{e66}) as a sum of
powers of $n$,
\begin{equation}
\frac{\Gamma(2j+1+n)}{n!}=\sum_{\beta=0}^{2j}(-1)^{2j+\beta}S_{2j}^{(\beta)}
\sum_{\sigma=0}^{\beta}\binom{\beta}{\sigma}\,n^{\sigma},
\label{e67}
\end{equation}
and then the zeta regularization (\ref{e59}) gives
\begin{widetext}
\begin{eqnarray}
\mu_k^{(j)} &=& (-1)^k\left\{ \right.\left(j+1/2\right)^{2k}+\frac{2^{2j+1}}{
\Gamma(2j+1)}\sum_{\alpha=0}^{2k-1}\binom{2k}{\alpha}\,\left(j+1/2\right)^{
\alpha}\nonumber\\
&& \times\sum_{\beta=0}^{2j}(-1)^{2j+\beta}S_{2j,\beta}^{1}\sum_{\sigma=0}^{
\beta}\binom{\beta}{\sigma}\left(2^{2k+\sigma-\alpha+1}-1\right)\zeta(-2k+\alpha
-\sigma)\left.\right\}.
\label{e68}
\end{eqnarray}
\end{widetext}
Equation (\ref{e68}) reproduces the particular values of ${\cal R}_k (j)=4^k
\mu_k^{(j)}/(2j+1)$ given in this paper for $j=0$ and $j=1/2$. The special
case ${\cal R}_k(-1/2)$ is obtained by first dividing (\ref{e68}) by $(2j+2)(2j+1)$
and then substituting the value $j=-1/2$ in the result (\ref{e53}).

For $(2j+1)<0$ the series over $n$ in (\ref{e66}) becomes finite and
regularization is not necessary. The result is 
\begin{eqnarray}
\mu_k^{(j)} &=& (-1)^k 2^{2j+1}\sum_{\alpha=0}^{2k}\binom{2k}{\alpha}\,\left(
j+1/2\right)^\alpha\nonumber\\
&&\times\Gamma(-2j)\sum_{n=0}^{-2j-1}\frac{n^{2k-\alpha}}{\Gamma (-2j-n)\,n!},
\label{e69}
\end{eqnarray}
which gives the same values for ${\cal R}_{n}(-1),\,{\cal R}_{n}(-5/2)$ derived
earlier in this paper.

The other way is to evaluate the moments of the polynomials $\cQ_k^{(j)}
(x)$ using their relation with the Meixner-Pollaczek polynomials (\ref{e61}).
Using the result \cite{S14} 
\begin{equation}
\int_{-\infty}^\infty dx\,e^{-(\pi-2\phi)x}|\Gamma(ix+a)|^2=\pi\,\Gamma\left(2a
\right)\left(2\sin\phi\right)^{-2a},
\label{e70}
\end{equation}
differentiating the above formula $n$ times, and dividing both sides by $2^n$,
we have
\begin{eqnarray}
&&\int_{-\infty}^\infty dx\,x^n e^{-(\pi-2\phi)x}|\Gamma(ix+a)|^2\nonumber\\
&&\quad=2^{-n}\pi\,\Gamma\left( 2a\right)\,\frac{d^n}{d\phi^n}(2\sin\phi)^{-2a}.
\label{e71}
\end{eqnarray}
From (\ref{e71}) with $\phi=\pi/2$ and $a=j+1/2$ and using (\ref{e61}) and
(\ref{e63}) we obtain the moments of $\cQ_k^{(j)}(x)$:
\begin{equation}
\mu_{2n}=2^{-2n-1}\frac{d^n}{d\phi^n}(2\sin\phi)^{-2j-1}\Big|_{\phi=\pi/2}.
\label{e72}
\end{equation}
This result is equivalent to (\ref{e68}) and (\ref{e69}).

Favard's theorem \cite{S15} states that if a sequence of monic polynomials
satisfies a three-term recursion relation of the form $xP_n(x)=P_{n+1}(x)+a_nP_n
(x)+b_nP_{n-1}(x)$ with $a_n,b_n\in\mathbb{R}$ and $b_n >0$, then there exists
a positive measure respect to which the polynomials are orthogonal on
$\mathbb{R}$. In our case, because the recursion relation for the polynomials
$\cQ_n^{(j)}(x)$ is $x\cQ_n^{(j)}(x)=\cQ_{n+1}^{(j)}(x)+n(n+2j)\cQ_{n-1}^{(j)}
(x)$, they are orthogonal with respect to a positive measure for $j\geq -1/2$, in
agreement with the results in (\ref{e52}) -- (\ref{e54}). In Ref.~\cite{S16} an
extension of the Meixner-Pollaczek polynomials for $\lambda\leq0$ is studied and
a nonstandard inner product is defined with respect to which they are
orthogonal.

\section{Conclusions}
\label{s5}

In this paper we have investigated the question of whether the minimal solution
for the time operator can be significantly simplified by adding homogeneous
solutions. To answer this question we had to make sense of and evaluate infinite
sums of singular operators, and to do this we calculated matrix elements of
these sums and then used zeta summation to sum the series. The matrix elements
give rise to remarkable Meixner polynomials having interesting mathematical
properties in which Euler and Bernoulli numbers repeatedly appear.

Our principal conclusion is that the matrix elements of the minimal solution
$\langle n-2j|\Theta_{\rm min}|n\rangle$ for $j=1,2\ldots$ and $n\geq2j$ {\it
cannot all be canceled} by adding one or more homogeneous solutions $\Theta_{\rm
hom}^{(\gamma)}$ for $\gamma=0,1,2, \ldots$ to the minimal solution.
Specifically, for the minimal solution the matrix elements $\langle n-4j+2|
\Theta_{\rm min}|n\rangle$ for $j=1,2\ldots$ and $n \geq 4j-2$ {\it can be
canceled} by adding to $\Theta_{\rm min}$ the homogeneous solution $\Theta_{\rm
hom}^{(0)}$ with the choice of the parameter $a_0^{(0)}=1/2$. However, the
homogeneous solutions $\Theta_{\rm hom}^{(\gamma)}$ for $\gamma\geq 1$ cannot
cancel {\it all} the matrix elements of the minimal solution $\langle n-4j+2|
\Theta_{\rm min}|n \rangle$ for $j=1,2\ldots$ and $n\geq 4j-2$.

To elaborate, we have shown that for a finite value of $\gamma$, the number of
nonzero columns of the operator $\Theta_{\rm hom}^{(\gamma)}$ is less than the
number of the nonzero columns of $\Theta_{\rm min}$. This means that it is
impossible to cancel all the matrix elements $\langle n-4j+2|\Theta_{\rm min}|
n\rangle$. As $\gamma\to\infty$, the number of nonzero elements in the matrix
representation of $\Theta_{\rm hom}^{(\gamma)}$ becomes equal to the number of
nonzero elements in the matrix representation of $\Theta_{\rm min}$. This
means that we can then consider the operator  
\begin{equation}
\Theta=\Theta_{\rm min}+\sum_{\gamma=0}^\infty \Theta_{\rm hom}^{(\gamma)}.
\label{e73}
\end{equation} 
The matrix elements in the first nonzero columns above and below the principal
diagonals ($j=1$) are
\begin{eqnarray}
\langle n-2|\Theta|n\rangle &=& -\frac{i}{2}\sqrt{\frac{n}{n-1}}(1-2a_0^{(0)})
\nonumber\\
&& \!\!\!\!\!\!\!\!\!\!\!\!\!\!\!\!\!\!\!\!\!\!\!\!\!\!\!\!\!\!\!\!\!\!\!\!\!
-\frac{i}{2}\sqrt{n(n-1)}\sum_{\sigma=0}^\infty a_0^{(\sigma+1)}\frac{(2\sigma
+2)!}{(\sigma+2)!2^\sigma\sigma!}\cP^{(1)}_\sigma(n).
\label{e74}
\end{eqnarray}
If we choose
\begin{equation}
a_0^{(\sigma)}=\frac{(\sigma+1)!}{(2\sigma)!}(2t)^{\sigma-1},
\label{e75}
\end{equation}
with $t$ to be determined, the matrix elements become
\begin{eqnarray}
\langle
n-2|\Theta|n\rangle &=& -\frac{(t-1)i}{2t}\sqrt{\frac{n}{n-1}}\nonumber\\
&& +4i\sqrt{n(n-1)}\frac{(t+2)^{n-2}}{(t-2)^{n+1}},
\label{e76}
\end{eqnarray}
where we have used the generating function of the polynomials $\cP^{(1)
}_\sigma(n)$ to sum the series in (\ref{e74}). The values of $t$ that make the
matrix elements (\ref{e76}) vanish are the roots of the equation
\begin{equation}
t(t-2)^{n+1}-8(n-1)t-(t-2)^{n+1}=0\quad(n\geq2),
\label{e77}
\end{equation}
with $t\neq 0,2$.

This demonstrates that to cancel all the elements in a fixed diagonal in the
matrix representation $\langle n-4j+2|\Theta|n\rangle$, the choice of the
parameter $a_0^{(\sigma)}$ must be different for each $n$. In general, the
value of $a_0^{(\sigma)}$ that cancels the matrix elements$\langle n-4j+2|\Theta
|n\rangle$ depends on both $j$ and $n$.

\acknowledgments
CMB thanks the U.S.~Department of Energy for financial support. MG thanks
G.~Landolfi for useful discussions, the Physics Department of Washington
University in St. Louis for its hospitality during the preparation of this
work, and the University of Salento and INFN (Lecce) for financial support.

\appendix
\setcounter{equation}{0}
\def\theequation{A\arabic{equation}}

\section{Calculation of Matrix Elements}
\label{a1}

In this Appendix we show how to calculate the matrix elements in (\ref{e31}).
Let us consider the operators
\begin{equation}
I_k=\left(\frac{1}{q}\frac{d}{dq}\right)^k,\quad
J_k=\left(\frac{d}{dq}\frac{1}{q}\right)^k\quad (k=1,2,\ldots).
\label{ea1}
\end{equation}
Here, $I_{k}$ acts on even eigenstates according to
\begin{widetext}
\begin{equation}
I_k\psi_n(q)=\sum_{\alpha=1}^k\sum_{\beta=0}^{k-\alpha}\frac{\alpha}{\alpha+2
\beta}\binom{k}{\alpha+\beta}R_{n,\alpha,\beta}\,
q^{-2\beta-\alpha}H_{n-\alpha}(q)e^{-x^2/2}-H_n(q)\,e^{-x^2/2},
\label{ea2}
\end{equation}
where $H_n$ are the Hermite polynomials. Similarly, $J_{k}$ acts on odd
eigenstates and gives the result
\begin{equation}
J_k\psi_n(q)=\sum_{\alpha=0}^k\sum_{\beta=0}^{k-\alpha}\binom{k}{\alpha+\beta}
R_{n,\alpha,\beta}q^{-2\beta-\alpha}H_{n-\alpha}e^{-x^2/2}.
\label{ea3}
\end{equation}
\end{widetext}
The numerical coefficients $R_{n,\alpha,\beta}$ are given by
\begin{equation}
R_{n,\alpha,\beta}=\frac{(-1)^{\alpha+1}2^{\alpha-\beta}(\alpha+2\beta)!n!}
{\pi^{1/4}\sqrt{2^n n!}\,\alpha!\beta!(n-\alpha)!}.
\label{ea4}
\end{equation}

The Hermite polynomials have the explicit form
\begin{eqnarray}
H_{n-\alpha}(q)&=&\sum_{l=0}^{\lfloor(n-\alpha)/2\rfloor}q^{n-\alpha-2l}
C_{n,\alpha,l},\nonumber\\
C_{n,\alpha,l}&=&(-1)^l 2^{n-\alpha-2l}\frac{(n-\alpha)!}{l!(n-\alpha-2l)!}.
\label{ea5}
\end{eqnarray}
Thus, we can evaluate the integral
\begin{eqnarray}
S_{n,m,\nu}&=&\int_0^\infty dq\,e^{-q^2}q^{n-2\nu}H_m(q)\nonumber\\
&=&\begin{cases}
\frac{2^{m-1}(n/2-\nu)!\Gamma(n/2-\nu+1/2)}{(n/2-\nu-m/2)!}&(m\leq n-2\nu),\\
0 & \text{otherwise}.
\end{cases}
\nonumber
\end{eqnarray}

The matrix elements $\langle m|I_k|n\rangle$ and $\langle m|J_k|n\rangle$ can
then be written in the form
\begin{eqnarray}
\langle m|I_k|n\rangle &
=&\sum_{\alpha=1}^{\alpha_{\rm max}}\sum_{\beta=0}^{\beta_{\rm
max}}\sum_{\ell=0}^{\ell_{\rm max}}\frac{\alpha}{\alpha+2\beta}\binom{2k+1}
{\alpha+\beta}\,U_{n,m,\alpha,\ell,\beta},\nonumber \\
\langle m|J_k|n\rangle &
=&\sum_{\alpha=0}^{\alpha_{\rm max}}\sum_{\beta=0}^{\beta_{\rm max}}
\sum_{\ell=0}^{\ell_{\rm max}}\binom{2k+1}{\alpha+\beta}\,U_{n,m,\alpha,\ell,
\beta},
\label{ea6}
\end{eqnarray}
where $\alpha_{\rm max}=(n-m)/2$, $\beta_{\rm max}=(n-m)/2-\alpha$,
$\ell_{\rm max}=(n-m)/2-\alpha-\beta$, and
\begin{equation}
U_{n,m,\alpha,\ell,\beta}=\frac{1}{\pi^{1/4}\sqrt{2^m
m!}}\,R_{n,\alpha,\beta}\,S_{m,n,\alpha+\ell+\beta}\,C_{n,\alpha,\ell}.
\label{ea7}
\end{equation}
The coefficients $U$ have the form
\begin{widetext}
\begin{eqnarray}
U_{2n+1,2m+1,\alpha,\ell,\beta}
&=&(-1)^{\alpha+\ell+1}2^{2\alpha+\beta+m-n}\sqrt{\frac{(2n+1)!}{(2m+1)!}}
\frac{(\alpha+2\beta)!(2n-2\alpha-2\beta-2\ell+1)!}{\beta!\ell!\alpha!
(n-m-\alpha-\beta-\ell)!(2n-\alpha-2\ell+1)!},\nonumber\\
U_{2n,2m,\alpha,\ell,\beta}
&=&(-1)^{\alpha+\ell+1}2^{2\alpha+\beta+m-n}\sqrt{\frac{(2n)!}{(2m)!}}
\frac{(\alpha+2\beta)!(2n-2\alpha-2\beta-2\ell)!}{\beta!\ell!\alpha!
(n-m-\alpha-\beta-\ell)!(2n-\alpha-2\ell)!}.\nonumber
\end{eqnarray}

\end{widetext}
With the change of variables $n=2n'$ and $m=2n'-2j$ for $j=1,2,...n'$, we
make the replacement
\begin{equation}
\tilde U_{n,j,\alpha,\ell,\beta}=U_{2n',2n'-2j,\alpha,\ell,\beta},
\label{ea8}
\end{equation}
and (\ref{ea6}) becomes
\begin{eqnarray}
\langle m|I_k|n\rangle & =&\sum_{\alpha=1}^{j}\ \sum_{\beta=0}^{j-\alpha}\
\sum_{\ell=0}^{j-\alpha-\beta}\frac{\alpha}{\alpha+2\beta}\binom{2k+1}
{\alpha+\beta}\,\tilde{U}_{n,j,\alpha,\ell,\beta},\nonumber\\
\langle m|J_k|n\rangle & =&\sum_{\alpha=0}^{j}\ \sum_{\beta=0}^{j-\alpha}\
\sum_{\ell=0}^{j-\alpha-\beta}\binom{2k+1}{\alpha+\beta}\
\tilde{U}_{n,j,\alpha,\ell,\beta}.
\label{ea9}
\end{eqnarray}
Then after the second change of variables, $\alpha=\rho-\sigma$, $\beta=\sigma$,
$\tilde{\tilde{U}}_{n,j,\rho,\ell,\sigma}=\tilde{U}_{n,j,\rho-\sigma,\ell,
\sigma}$, we obtain
\begin{eqnarray}
\langle m|I_k|n\rangle & =&\sum_{\rho=1}^{j}\ \sum_{\sigma=0}^{\rho-1}\
\sum_{\ell=0}^{j-\rho}\frac{\rho-\sigma}{\rho+\sigma}\binom{2k+1}{\rho}\,
\tilde{\tilde{U}}_{n,j,\rho,\ell,\sigma},\nonumber \\
\langle m|J_k|n\rangle & =&\sum_{\rho=1}^{j}\ \sum_{\sigma=0}^{\rho-1}\
\sum_{\ell=0}^{j-\rho}\binom{2k+1}{\rho}\
\tilde{\tilde{U}}_{n,j,\rho,\ell,\sigma}.
\label{ea10}
\end{eqnarray}

Next, we use the formulas
\begin{eqnarray}
\langle 2n-2j|\Theta_{\rm min}|2n\rangle &=&\nonumber\\
&&
\!\!\!\!\!\!\!\!\!\!\!\!\!\!\!\!\!\!\!\!\!\!\!\!\!\!\!\!\!\!\!\!\!\!\!\!\!\!\!
\!\!\!\!\!\!\!
-\frac{i}{2}\sum_{k=0}^\infty\frac{1}{2k+1}\langle 2n-2j|I_{2k+1}|2n\rangle,
\label{ea11}
\end{eqnarray}
\begin{eqnarray}
\langle
2n+1-2j|\Theta_{\rm min}|2n+1\rangle &=&\nonumber\\
&& \!\!\!\!\!\!\!\!\!\!\!\!\!\!\!\!\!\!\!\!\!\!\!\!\!\!\!\!\!\!\!\!\!\!\!\!\!\!
\!\!\!\!\!\!\!\!\!\!\!\!\!\!\!\!\!\!\!\!\!\!\!\!\!\!\!\!\!\!\!\!\!\!\!\!\!\!\!
-\frac{i}{2}\sum_{k=0}^{\infty}\frac{1}{2k+1}\langle 2n+1-2j|J_{2k+1}|2n+1
\rangle,
\label{ea12}
\end{eqnarray}
to rewrite the even-even and odd-odd matrix elements
\begin{eqnarray}
\langle2n-2j|\Theta_{\rm min}|2n\rangle &=& \nonumber\\
&& \!\!\!\!\!\!\!\!\!\!\!\!\!\!\!\!\!\!\!\!\!\!\!\!\!\!\!\!\!\!\!\!\!\!\!\!\!
\!\!\!\!\!\!\!\!\!\!\!\!\!\!\!\!\!\!\!\!\!\!\!\!\!\!\!\!\!\!\!\!\!\!\!\!\!\!\!
\!\!\!\!\!\!\!\!\!\!\!\!\!\!\! {\cal R}_{n,j}^e\sum_{\rho=1}^j\frac{2^\rho}{(
\rho-1)!(j-\rho)!}\sum_{k=0}^\infty\frac{1}{2k+1}\binom{2k+1}{\delta},
\nonumber \\
\langle 2n+1-2j|\Theta_{\rm min}|2n+1\rangle &=& \nonumber\\
&& \!\!\!\!\!\!\!\!\!\!\!\!\!\!\!\!\!\!\!\!\!\!\!\!\!\!\!\!\!\!\!\!\!\!\!\!\!
\!\!\!\!\!\!\!\!\!\!\!\!\!\!\!\!\!\!\!\!\!\!\!\!\!\!\!\!\!\!\!\!\!\!\!\!\!\!\!
\!\!\!\!\!\!\!\!\!\!\!\!\!\!\! {\cal R}_{n,j}^o\sum_{\delta=1}^j\frac{2^\delta}
{(\delta-1)!(j-\delta)!}\sum_{k=0}^\infty\frac{1}{2k+1}\binom{2k+1}{\rho},
\label{ea13}
\end{eqnarray}
where for $n=1,2,\ldots$ and $j=1,2,\ldots n$
\begin{eqnarray}
\mathcal{R}_{n,j}^e &=& i (-1)^j 2^{j/2-1}(j-1)!
\sqrt{\frac{n!(2n-2j-1)!!}{(n-j)!(2n-1)!!}},\nonumber\\
\mathcal{R}_{n,j}^o &=& i (-1)^{j+1}2^{j/2-1}(j-1)!
\sqrt{\frac{n!(2n-2j+1)!!}{(n-j)!(2n+1)!!}}.
\nonumber
\end{eqnarray}

Now an apparent problem arises: The series over $k$ in (\ref{ea13}) is
divergent. To overcome this problem we resort to zeta summation to evaluate the
divergent series. Zeta summation is conventionally used to evaluate a divergent
sum over modes. (For example, it is used in the calculation of the Casimir force
\cite{S8}.) The idea behind zeta summation is to extend $\zeta(s)=
\sum_{k=1}^\infty k^{-s}$, which is valid for ${\rm Re}\,s>1$, to negative
values of $s$. To do so, one must use the functional equation for the Riemann
zeta function
\begin{equation}
\zeta(1-s) =2(2\pi)^{-s}\Gamma(s)\cos\left(\frac{\pi s}{2}\right)\zeta(s).
\label{ea14}
\end{equation}
We can thus express the divergent series $\sum_{k=1}^\infty a(k)$ as
$\lim_{s\to0}\sum_{k=1}^\infty a(k)k^{-s}$. This gives results such as
$$\zeta(0)=\sum_{k=1}^\infty k^{0}=-\frac{1}{2},
\quad\zeta(-1)=\sum_{k=1}^\infty k=-\frac{1}{12}.$$

In our case the sum 
\begin{equation}
Z(\rho)=\sum_{k=0}^\infty\frac{1}{2k+1}\binom{2k+1}{\rho}
\label{ea15}
\end{equation}
becomes
\begin{equation}
Z(\rho)=\delta_{\rho,1}+\lim_{s\to 0}\sum_{k=1}^{\infty}\binom{2k+1}
{\rho}\frac{k^{-s}}{2k+1}.
\label{ea16}
\end{equation}
It is now straightforward to evaluate the sum over $k$ for generic values of the
parameter $\rho$. We first rewrite the sum by expressing the binomial
coefficient as a polynomial in the variable $k$
\begin{equation}
\binom{2k+1}{\rho}=\sum_{i=1}^{\rho}\frac{S_{\rho,i}^1}{\rho!}(2k+1)^i~
(\rho=1,2,3\ldots),
\label{ea17}
\end{equation}
where $S_{\rho,i}^1$ are the Stirling numbers of first kind. Then,
\begin{equation}
Z(\rho)=\delta_{\rho,1}+\sum_{i=1}^\rho\frac{S_{\rho,i}^1}{\rho!}
\sum_{r=0}^{i-1}\binom{i-1}{r}\ 2^r\zeta(-r).
\label{ea18}
\end{equation}

Because $\zeta(0)=-1/2$ , $\zeta(-2n)=0$ and $\zeta(1-2n)=-\cB_{2n}/(2n)$ for $n
=1,2,3\ldots$, (\ref{ea16}) is nonzero only for odd values of $r$ and for $r=0$.
Replacing $r\to2r-1$, we obtain the final results in (\ref{e32}) and
(\ref{e33}).


\begin{thebibliography}{99}
\bibitem{S6} Another simple relation is $T_{n,n}=P_n\left(T_{1,1}\right)$,
where $P_n$ is the $n$th Hahn polynomial; this was derived in C.~M.~Bender,
L.~R.~Mead, and S.~S.~Pinsky, J.~Math.~Phys.~{\bf 28}, 509 (1987).
\end{thebibliography}
\end{document}